\documentclass[12pt]{iopart}

\usepackage{graphicx}

\begin{document}

\title{Evaluating cumulative ascent: Mountain biking meets Mandelbrot}

\author{D. C. Rapaport}

\address{Department of Physics, Bar-Ilan University, Ramat-Gan 52900, Israel}

\ead{rapaport@mail.biu.ac.il}

\date{February 08, 2011}

\begin{abstract}

The problem of determining total distance ascended during a mountain bike trip
is addressed. Altitude measurements are obtained from GPS receivers utilizing
both GPS-based and barometric altitude data, with data averaging used to reduce
fluctuations. The estimation process is sensitive to the degree of averaging,
and is related to the well-known question of determining coastline length.
Barometric-based measurements prove more reliable, due to their insensitivity to
GPS altitude fluctuations.

\noindent{\it Keywords\/}: GPS, altitude measurement, Mandelbrot, fractals

\end{abstract}

\pacs{91.10.Fc, 05.45.Df}

\section{Introduction}

The typical mountain bike ride includes changes of altitude, and the effort
expended depends, in part, on the overall distance climbed. Examples of other
contributions to overall effort, some neither quantifiable nor reproducible,
include riding speed, terrain conditions and weather, each of which can vary
considerably during a ride. The total vertical ascent is an apparently simple
characteristic of a given route, so that it would be appropriate to assign it a
numerical value. It turns out that this is a more difficult task than
anticipated, with the measurement process itself not necessarily able to provide
a unique answer.

The problem is analogous to the well-known exercise of measuring the length of
the coastline of a country, with Great Britain the instance headlined by the
work of Richardson and Mandelbrot [1,2]. Length estimates can vary
substantially, depending on the scale of the map used, because smaller coastline
features appear as resolution is increased, an observation that leads to the
consideration of irregular forms in general. For the overachieving cyclist,
however, estimating cumulative ascent is an important practical matter. While
the late Benoit Mandelbrot is not reported to have encountered mountain biking
in person, the methodology spawned by adopting a fractal perspective when taking
the measure of geometrically irregular shapes can offer guidance in addressing
the question.

Modern technology, in the form of the Global Positioning System (GPS) satellite
constellation, and affordable, compact GPS receivers, provides the necessary
information. A GPS receiver acquires data from multiple satellites, which it
processes to supply the user with a relatively precise geographical location and
an altitude estimate. Location is expressed either as latitude and longitude or,
after conversion, as map grid coordinates, while altitude is measured relative
to a model of the earth's essentially ellipsoidal geoid. Positional accuracy
depends on the locations of the satellites offering the strongest signals, as
well as on local topography and other obstructions that might degrade signal
quality.

An alternative means of altitude measurement relies on the height variation of
atmospheric pressure. A barometric device, after suitable calibration, can be
used to measure height above a reference level, assuming no pressure changes due
to meteorological causes. Certain GPS receivers incorporate a barometric
altimeter that is continually recalibrated using GPS data, but which, due to a
relatively slow response, is far less susceptible to varying GPS signal quality
and thus provides more stable results.

The present paper describes the outcome of tests using both types of altitude
measurement. Different degrees of data averaging are applied to reduce the
effect of the measurement `noise'. It is apparent from the results that there is
no preferred length scale in the problem that could help determine the optimal
averaging, leading to the somewhat unexpected conclusion that there is really no
correct answer to the question, the same as reached when attempting to measure
the length of the British coastline.

\section{Methods}

Two GPS receivers were mounted on the same bicycle to ensure they followed a
common space-time trajectory. One was the Garmin Edge 205, a receiver intended
for cycling use that employs GPS-based altitude measurement, while the other was
the Garmin Dakota 20, a more general-purpose device that also incorporates a
barometric altimeter [3]. Each receiver records its position history at a
variable rate designed to allow reproduction of an accurate track, ideally of a
quality suitable for subsequent almost map-free navigation. The history consists
of a series of trackpoints, each specifying time of day, horizontal position
expressed as latitude and longitude, and altitude; it can be retrieved for
computer processing, in the case of the Edge using, e.g., the open source
GPSBabel [4] software, and for the Dakota by copying the GPX-format data file.
Trackpoints are converted to map coordinates with the appropriate transformation
functions [5] -- the relevant conversion for the present work being the
Cassini-Soldner projection -- borrowed from the GPSBabel source code; horizontal
distance measurements then follow immediately.

The hilly terrain conditions over the first of the closed routes that provided
data for analysis are typical for offroad cycling, with partial tree cover and
topography capable of degrading GPS signal quality; results from two other
trips, one involving similarly hilly conditions and the other over flatter
terrain, are discussed briefly towards the end of the paper. Good horizontal
positional accuracy is achieved over the 42 km (measured horizontally) track,
and the tracks from the two receivers are essentially identical when
superimposed on a 1:50000 scale map. While there are many sources of error,
ranging from suboptimal GPS signal reception to the analog-to-digital conversion
of the barometric readings, no additional information, e.g., details of the
proprietary algorithms used to process the raw GPS data, is readily available
that could be utilized in the analysis.

The methods used by the receivers for determining recording rate are not
specified, but the measured time-interval distribution for the Edge ($\sim$2280
points recorded) has a sharp peak at 6.5s and a small peak at 1s, with
essentially all intervals $<$ 13s (longer intervals correspond to a nominally
stationary receiver), while for the Dakota ($\sim$1600 points) there is a peak
at 12s and a weakly split peak in the range 1-4s with most intervals $<$ 20s;
the time intervals show no obvious speed dependence (those above 20s only appear
at speeds below 5 km/h). The corresponding distance-interval distribution for
the Edge peaks at 20m, while the Dakota has a peak at 8m and a weakly split peak
over 20-40m. The maximum distance interval increases with speed, and for any
given speed there is a broad spread of values; the maxima (excluding a few
outlying points) at the overall mean speed (12 km/h) are 40m for the Edge and
60m for the Dakota. Larger intervals might affect distance accuracy if there are
sharp changes in direction, although these become less likely at higher speeds
(and direction changes may even reduce the interval size, another unknown aspect
of receiver behavior). Variable intervals allow much longer track histories to
be recorded, compared to the use of a fixed and necessarily short interval, but
their use can complicate the altitude analysis.

\section{Results}

\begin{figure}
\begin{center}
\includegraphics[scale=1.1]{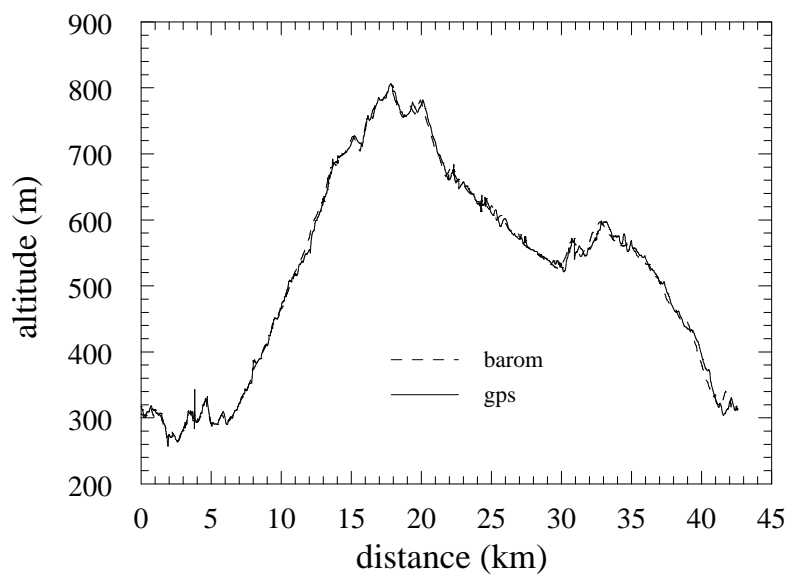}
\end{center}
\caption{\label{fig1} Altitude estimates vs distance obtained from the Garmin
Dakota (barometric data) and Edge (GPS data) receivers.}
\end{figure}

\begin{figure}
\begin{center}
\includegraphics[scale=1.1]{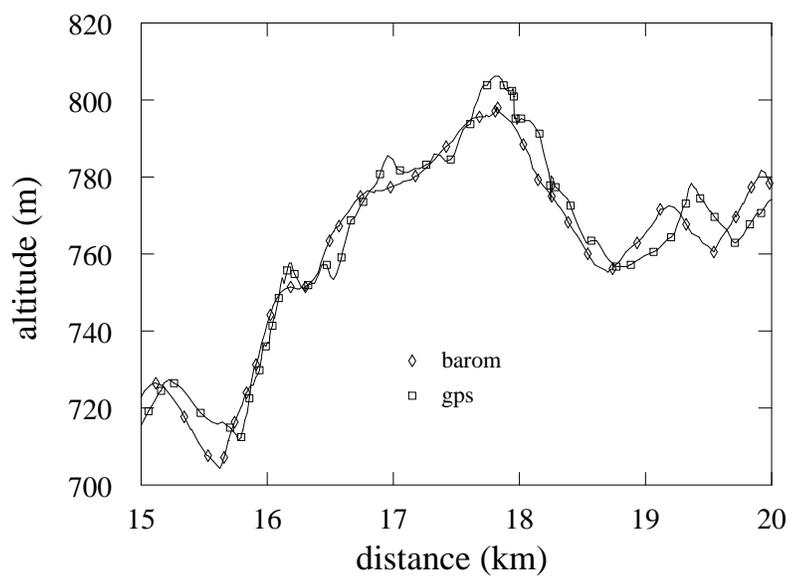}
\end{center}
\caption{\label{fig2} Enlarged portion of Fig.~\ref{fig1} showing the
differences in altitude data (symbols every 5th point).}
\end{figure}

Altitude data measured by the two GPS receivers over the duration of the first
trip are shown in Fig.~\ref{fig1}. The two sets of data overlap reasonably well,
after allowing for a 0.4 km (1\%) difference between the horizontal track length
estimates, that of the Edge being longer, with the difference probably due to
the fact that 40\% more trackpoints are recorded. However, the GPS-based
altitude data from the Edge is considerably more noisy than the barometric
Dakota; an example is shown in Fig.~\ref{fig2}. Altitude measurements based only
on the GPS signal are susceptible to spikes and other irregularities due to
momentary signal degradation, e.g., the prominent spike in Fig.~\ref{fig1} at 4
km associated with signal loss while in a tunnel under a highway; averaging will
be essential to reduce their otherwise serious impact on cumulative ascent
evaluation. The altitude varies between 260m and 800m over the route, so while
this 540m difference represents a lower bound for the total ascent, the question
is how much extra climbing is actually present in the data?

\begin{figure}
\begin{center}
\includegraphics[scale=1.1]{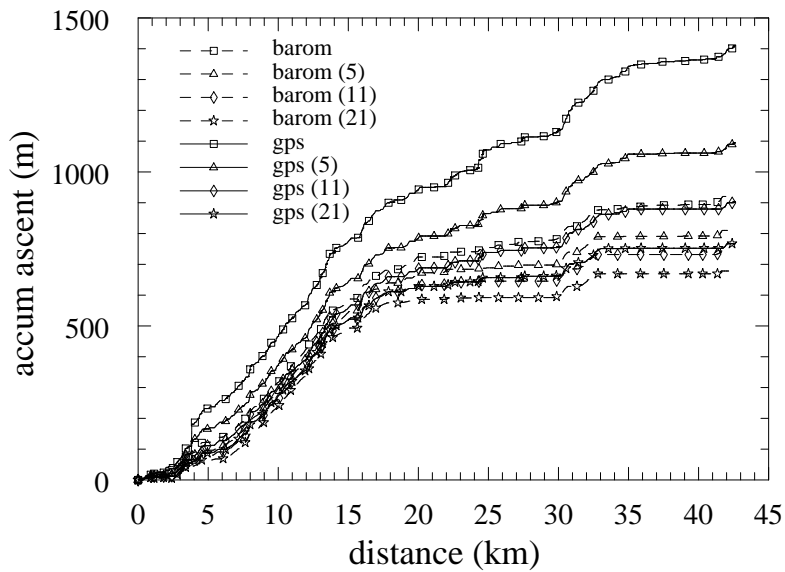}
\end{center}
\caption{\label{fig3} Accumulated ascent vs distance from the data of
Fig.~\ref{fig1}; the original data and averages over specified numbers of points
are shown (symbols every 50th point).}
\end{figure}

The spurious contributions of fluctuating altitude measurements can be reduced
by averaging. For the initial attempt at analysis, symmetric equally-weighted
averages are evaluated about each data point, where the number of points
included is regarded as a parameter. The cumulative ascent is then computed
along the track. Results based on averages over 5, 11 and 21 successive points,
as well as on the unaveraged data, are shown in Fig.~\ref{fig3}. The terminal
value of each curve represents an estimate of the total ascent. These are seen
to range from $\sim$680m to 1410m, and while the largest values are clearly
gross overestimates, and the smallest underestimates, the remaining values, all
seemingly plausible, are spread over a relatively broad $\sim$200m. For a given
degree of averaging, estimates based on GPS altitudes are all substantially
larger than the barometric results, consistent with the increased fluctuations
noted previously.

\begin{figure}
\begin{center}
\includegraphics[scale=1.1]{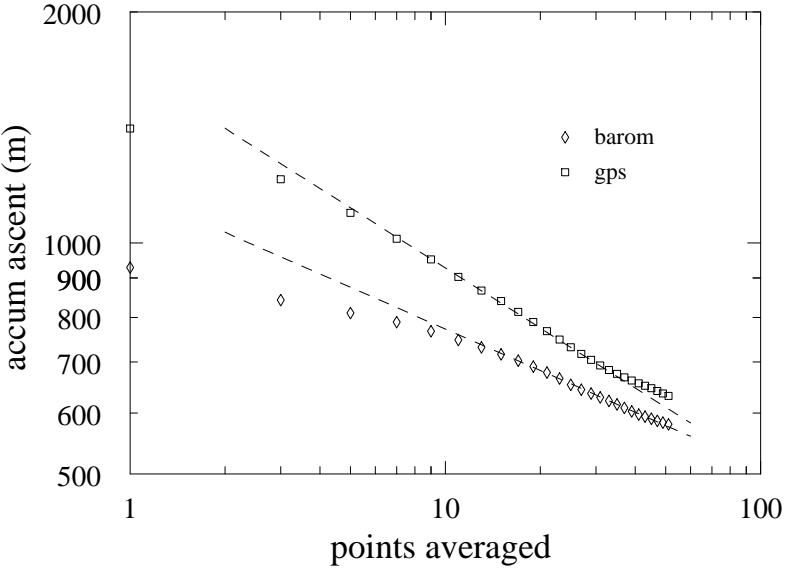}
\end{center}
\caption{\label{fig4} Log-log plots of total accumulated ascent based on
averaging over different numbers of points from the data of Fig.~\ref{fig1},
together with power-law fits (see text).}
\end{figure}

A systematic examination of the dependence of total cumulative ascent $C_a(s)$
on $s$, the number of points used in the averaging -- itself only a rough
measure since the distance between consecutive trackpoints is not fixed -- is
carried out with the aid of a log-log plot. This is a widely-used technique for
analyzing data from processes considered to be devoid of intrinsic size scales,
and the result appears in Fig.~\ref{fig4}. The absence of a plateau is a
signature of the scale-free nature of the problem. Reasonably good fits of the
power law expression [2] $C_a(s) = G s^{(1 - D)}$ to the plotted data, over
approximately a single decade, lead to exponents $D = 1.26$ and 1.18 for the GPS
and barometric altitudes respectively, but see [6,7] for a discussion of how
many decades are needed to establish `fractality'. Unfortunately, such an
outcome is of limited interest, given that the quantity $s$ does not correspond
directly, but only on average, to a physical distance.

\begin{figure}
\begin{center}
\includegraphics[scale=1.1]{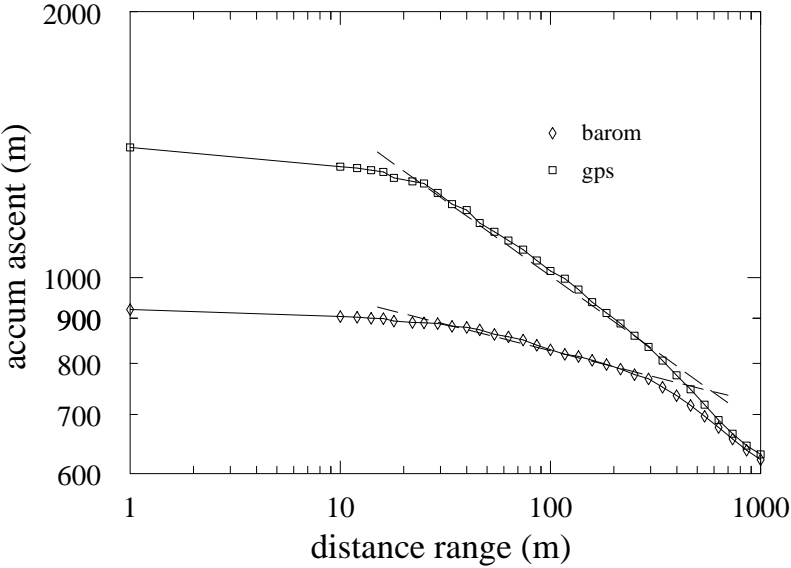}
\end{center}
\caption{\label{fig5} Log-log plots of total accumulated ascent based on
averaging using a distance window applied to the data of Fig.~\ref{fig1}, for
different distance ranges; power-law fits are included (see text).}
\end{figure}

A more meaningful approach to reducing the effects of fluctuations takes the
real separation of successive trackpoints into consideration, with a sliding
distance window used to select the series of neighboring points that contribute
to the averaged altitude at each data point. Although additional computation is
involved, and the method is unsuitable for real-time usage since the averaging
is symmetric, it corresponds to the methods used for coastline measurement [2].
Fig.~\ref{fig5} shows a log-log plot of the results obtained using windows
spanning different distance ranges. Power-law fits over the near-linear 30-300m
range yield exponents $D = 1.17$ and 1.06 for the GPS and barometric altitudes,
with the larger value, as before, a consequence of the GPS-based altitude
fluctuations.

In view of typical terrain conditions, the $\sim$20-50m window range is likely
to be the most relevant. Over this range the barometric data yields a spread of
ascent values between 860m and 890m, and in the absence of averaging the value
increases slightly to 920m, so the uncertainty amounts to $\sim$4\%. The much
larger GPS-based values are seen to need averaging over almost 200m on average
before reaching even the unaveraged barometric estimate, with a 200m wider data
window required for similar estimates at intermediate ranges, e.g., 300m vs
100m; this essentially fictitious contribution to the cumulative ascent should
grow linearly with the time spent in motion. At widths beyond 700m both sets of
estimates are essentially the same, suggesting that GPS-based averages over such
large distance ranges are no longer sensitive to the measurement fluctuations;
however, excessive averaging will erase real topographical features, such as
intermediate hilltops, leading to an underestimated total ascent.

\begin{figure}
\begin{center}
\includegraphics[scale=1.1]{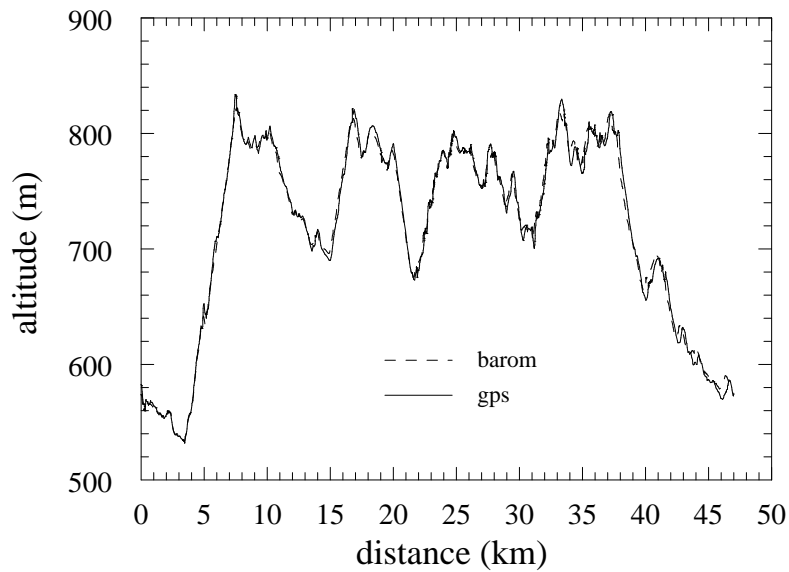}
\end{center}
\caption{\label{fig6} Altitude estimates vs distance from the two receivers over
the second hilly route.}
\end{figure}

\begin{figure}
\begin{center}
\includegraphics[scale=1.1]{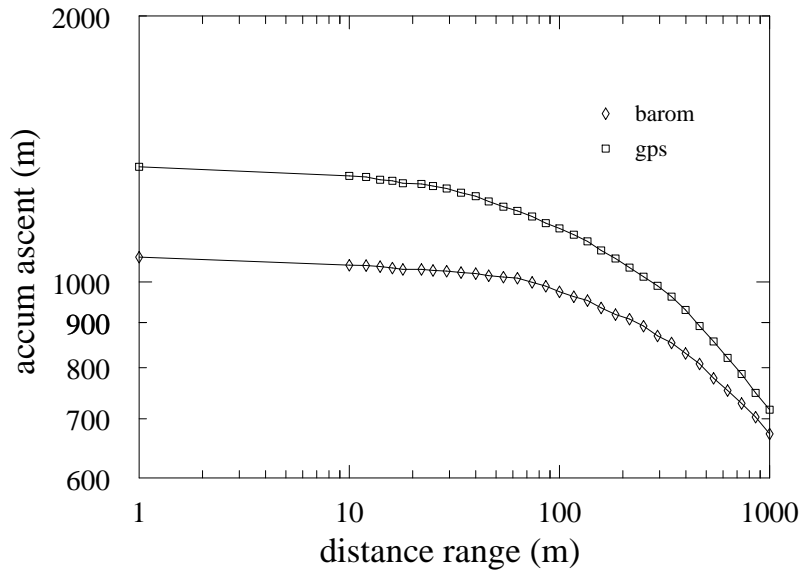}
\end{center}
\caption{\label{fig7} Total accumulated ascent based on averaging using a
distance window applied to Fig.~\ref{fig6}.}
\end{figure}

Similar analysis was carried out for the track whose altitude measurements
appear in Fig.~\ref{fig6} (track length 47 km, with 1640 and 2210 trackpoints
recorded); although difficult to see at this scale, the altitude plots are
subject to the same kinds of small differences as in Fig.~\ref{fig1}.
Accumulated ascent estimates using distance-based averaging are shown in
Fig.~\ref{fig7}. The barometric measurements using 20-50m averaging windows are
in the range $\sim$1010-1030m, increasing to 1070m without averaging. The
GPS-based results overestimate ascent as before, with averaging over at least
$\sim$200m again needed to recover the barometric estimates. (The near-linear
range available for a power law fit is reduced to less than a decade, with fits
over the 50-200m range yielding $D = 1.13$ and 1.10, values within a few percent
of the previous track; note that coastline measurements also do not indicate $D$
to be universal.)

\begin{figure}
\begin{center}
\includegraphics[scale=1.1]{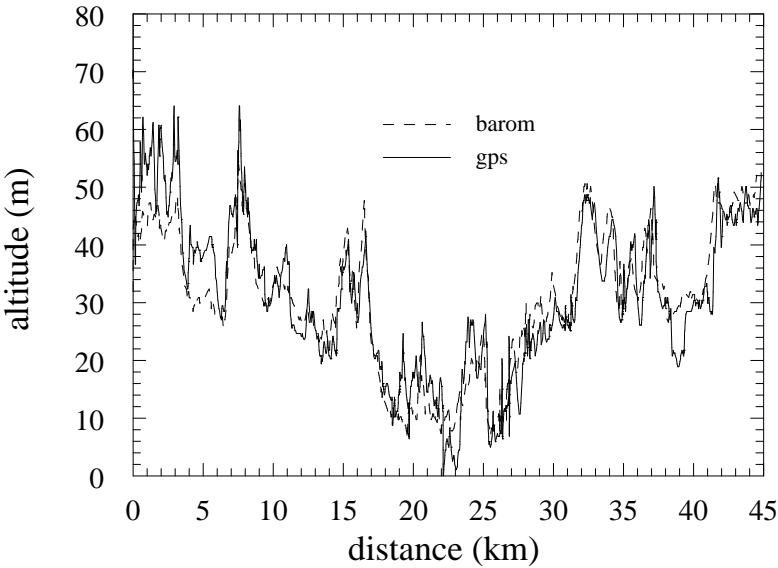}
\end{center}
\caption{\label{fig8} Altitude estimates vs distance for a relatively flat
route.}
\end{figure}

\begin{figure}
\begin{center}
\includegraphics[scale=1.1]{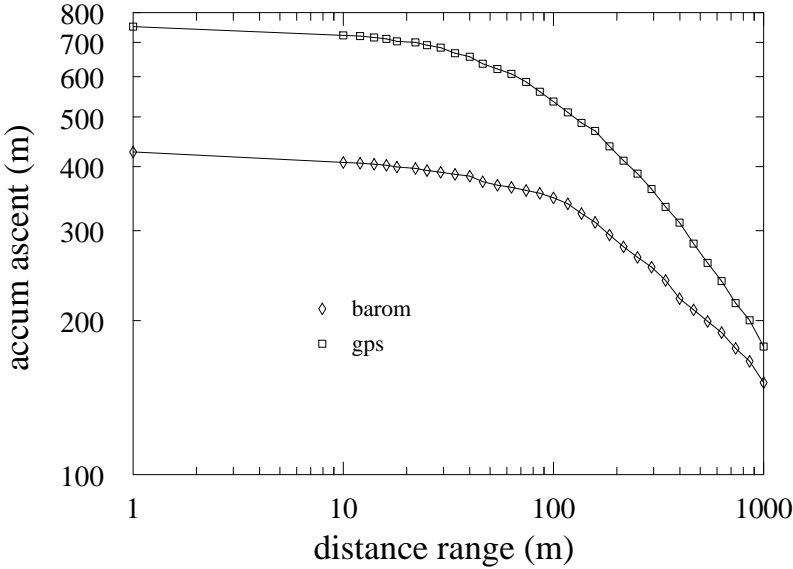}
\end{center}
\caption{\label{fig9} Total accumulated ascent from the data of
Fig.~\ref{fig8}.}
\end{figure}

The final example involves a track free of significant altitude changes (length
44 km, 1240 and 1850 trackpoints recorded). Altitude measurements are shown in
Fig.~\ref{fig8}, where the GPS-based results are again subject to larger
fluctuations than the barometric values. The fact that the fluctuations appear
more prominent is due to the greatly reduced altitude range, now only 60m. While
in the previous cases it might have been possible to make rough visual estimates
of the total ascent, here the result is less apparent.

The accumulated ascent estimates using distance-based averaging are shown in
Fig.~\ref{fig9}. For this track, the GPS-based results overestimate ascent by a
substantial factor, rendering them useless in practice. The barometric
measurements are more reasonable, with the 20-50m averaging range again
providing a good basis for estimation. The total ascent is in the range
$\sim$370-400m, a similar spread as before (but now amounting to 8\%), while
omitting the averaging adds another 30m; larger windows again tend to erase real
terrain features. (In the case of the GPS-based altitudes, exponent estimation
is not possible, whereas the barometric altitudes apparently exhibit two
distinct subranges below and above the 100m value, with exponents $D = 1.09$ and
1.35. Averaging based on fixed numbers of data points leads to $D = 1.63$ and
1.44, values larger than for the first track that reflect the enhanced
contribution of short-range fluctuations in the data.)

Determining the extent to which observations of this kind are generally
applicable requires more extensive experimentation covering a wider selection of
topographical conditions, although the trend seems clear. It should also be
pointed out that while averaging tends to eliminate small-scale features that
might be little more than `humps' along the trail, not all such humps are equal;
a small ascent followed by the corresponding descent might go almost unnoticed
when embedded in a horizontal or downhill segment, simply due to momentum, but
if superimposed on an already significant uphill grade its presence will
certainly be felt. Thus, even a well-defined measure of ascent would be unable
to characterize completely its contribution to the overall effort.

\section{Conclusion}

The outcome of this analysis, which is not offered as a systematic study of the
problem but merely as a limited set of observations, is the conclusion that
estimating cumulative ascent, an important measure of effort expended by a
cross-country cyclist, is an ill-defined task. At best, a range of estimates can
be obtained, hopefully one that is comparatively narrow. Furthermore, the use of
barometric altitude data appears essential for meaningful results; GPS-based
altitude data, without a substantial degree of averaging, leads to
overestimation. The question as to which particular estimate is the `correct'
one, just as when measuring the length of the coast of Britain (or almost any
other country, except Nauru), has no unique answer; an approach to the data
analysis based on that used for studying fractal-like phenomena helps clarify
the situation. While the coastline problem might be more of an intellectual
curiosity, cumulative ascent -- at least to some -- is of considerable practical
importance.

\end{document}